\documentclass[12pt]{revtex4}
\usepackage{epsfig}
\usepackage{amssymb}
\usepackage{amsmath}
\usepackage{bm}
\newcommand\xb{\mathbf{x}}
\newcommand\bb{\mathbf{b}}
\newcommand\vb{\mathbf{v}}
\newcommand\Ab{\mathbf{A}}
\newcommand\Eb{\mathbf{E}}
\newcommand\Bb{\mathbf{B}}
\newcommand\Rb{\mathbf{R}}
\newcommand\ub{\mathbf{u}}

\newcommand\Mb{\mathbf{M}}

\newcommand\gb{{\mathbf g}} 

\newcommand\db{\mathbf{d}}
\newcommand\rhob{\bm{\rho}}
\newcommand\thetab{\bm{\theta}}

\newcommand\ddt{\frac{\partial}{\partial t}}

\newcommand\daone{\delta_1 \mathbf{A}}
\newcommand\daonet{\delta_1 \tilde{\mathbf{A}}}
\newcommand\datwo{\delta_2 \mathbf{A}}
\newcommand\datwot{\delta_2 \tilde{\mathbf{A}}}
\newcommand\dpone{\delta_1 \phi}

\newcommand\dptwo{\delta_2 \phi}
\newcommand\dptwot{\delta_2 \tilde{\phi}}

\newcommand\dsdtheta{\frac{\partial S}{\partial \theta}}

\newcommand\ddtheta{\frac{\partial}{\partial \theta}}
\newcommand\ddmu{\frac{\partial}{\partial \mu}}
\newcommand\dMdtz{\frac{d M}{d t_0}}
\newcommand\dMdtheta{\frac{\partial M}{\partial \theta}}
\newcommand\dMdmu{\frac{\partial M}{\partial \mu}}
\newcommand\drhodtz{\frac{d \rho}{d t_0}}

\newcommand\numberthis{\addtocounter{equation}{1}\tag{\theequation}}

\begin{document}

\title{A very general electromagnetic gyrokinetic formalism.}
\date{\today}
\author{B.F. McMillan}
\affiliation{Centre for Fusion, Space and Astrophysics, Department of Physics, University of Warwick, CV4 7AL, Coventry UK}
\author{A. Sharma}
\affiliation{Centre for Fusion, Space and Astrophysics, Department of Physics, University of Warwick, CV4 7AL, Coventry UK}

\begin{abstract}
  We derive a gyrokinetic formalism which is very generally valid: the ordering allows both large inhomogeneities in plasma flow and
  magnetic field at long wavelength, like typical drift-kinetic theories, as well as fluctuations at the gyro-scale. The underlying
  approach is to order the vorticity to be small, and to assert that the timescales in the local plasma frame are long compared to the gyrofrequency.
  Unlike most other derivations, we do not treat the long and short wavelength components of the fluctuating fields separately; the
  single-field description defines the particle motion and their interaction with the electromagnetic field at small-scale, the
  system-scale, and intermediate length scales in a unified fashion. As in earlier literature, the work consists of identifying a
  coordinate system where the gyroangle-dependent terms are small, and using a near-unity transform to systematically
  find a set of coordinates where the gyroangle dependence vanishes.
\end{abstract}

\maketitle

We derive a gyrokinetic Lagrangian which is valid where the vorticity $ |\nabla \times (\Eb \times \Bb/B )| $
is small compared to the gyrofrequency $\Omega$, and the magnetic field scale length is long
compared to the gyroradius; we also require that time variation be slow in an appropriately chosen reference frame. 
This appears to be a minimum set of constraints on a gyrokinetic theory, and is substantially
more general than earlier approaches. It is the general-geometry electromagnetic extension of 
Ref \cite{Dimits_strongflow} (which is an electrostatic formalism with a homogeneous background magnetic field).
This approach also does not require a separate treatment of fluctuating and background components
of the magnetic field, unlike much of the previous literature. As a consequence, the `cross terms'
due to a combination of long- and short-wavelength variation, that
were ignored in earlier work (but derived in a more restrictive ordering in Ref. \cite{Parra_calvo}),
also appear naturally.

\section{ The guiding centre transform.}

The classical, non-relativistic Lagrangian of a particle in an electromagnetic field may be written
\begin{equation}
 \gamma = [ \Ab(\xb) + \vb ]. d\xb - \left[ \frac{1}{2} \vb^2 + \phi(\xb,t) \right] dt
\end{equation}
in dimensionless units where physical electromagnetic fields are recovered by multiplying by the mass
to charge ratio.

In a strongly magnetised plasma, the basic motion associated with this Lagrangian is a rapid
oscillation of the velocity and a helical trajectory in space. The rest of this paper is concerned with
the manipulation of this Lagrangian so that this fast gyration and
the slow dynamics referred to as drift motion may be treated independently.

We introduce a, for the moment arbitrary, velocity shift and
redefine $\vb$ as $\vb \rightarrow \vb + \ub$ so
\begin{equation}
 \gamma = [ \Ab(\xb) + \vb + \ub ]. d\xb - \left[\frac{1}{2} (\vb+\ub)^2 + \phi(\xb,t)\right] dt
\end{equation}

We will use the `local' ordering, where $\Ab,\Bb,\phi,\xb,\ub,\vb$ are taken to be of order $1$. Typical gyroradii
are then of order unity; this choice is fairly standard but some texts use a `global' ordering where the fields are
ordered $1/\epsilon$ and the gyroradius is ordered to be small.

The guiding centre transformation is written explicitly in terms of the quantities
$\Rb,\rhob$ and $U$. $\rhob$ is a point on the plane perpendicular to the local field direction 
$\hat{\bb}(\Rb,t)$. We rewrite $\vb$ in terms of a parallel velocity and the displacement $\rhob$
from the guiding centre, via:
\begin{equation}
 \vb = B(\Rb,t) \rhob \times \bb(\Rb,t) + U \bb(\Rb,t) + \ub(\Rb,\rhob,t)
\end{equation}
With the guiding centre $\Rb$ defined via
\begin{equation}
 \Rb = \xb - \rhob
\end{equation}

This transform is equivalent to that of Littlejohn (1983) only at lowest order
because the definition of the velocity vector uses directions at the guiding centre 
rather than the particle position; Littlejohn's approach simplifies the low-order, long-wavelength 
terms of the Lagrangian somewhat, but appears counterproductive when short-wavelength modes enter and 
we need to go one order higher. Later, when we perform the Lie transform, the constraint that 
oscillatory terms vanish tends to constrain the overall transform to be equivalent in regimes
where both are valid. 

We then follow the steps of Littlejohn (1983), with the crucial additional exception that instead of
using the Taylor series expansions of $\Ab$ and $\phi$, these are split as
\begin{equation}
\Ab(\xb,t) = \Ab(\Rb,t) + \rhob \cdot \nabla \Ab( \Rb,t) 
               + \delta_1 \Ab(\Rb,\rhob,t)
\end{equation}
and
\begin{equation}
\phi(\xb,t) = \phi(\Rb,t) + \rhob \cdot \nabla \phi(\Rb,t) + \delta_1 \phi(\Rb,\rhob,t).
\end{equation}
The assumed order $1$ variation of $A$ and $\phi$ gives the correct ordering for the zeroth and first derivative,
which are mostly due to long spatial variations, but the requirement that vorticity be small implies that the second derivatives are   
higher order. That is, the terms $\delta_1 \Ab$ and $ \delta_1 \phi$ are small.
The time derivatives of the electric and magnetic fields are also taken to be small.

Large amplitude system-scale features are mostly captured by the first derivative of the potentials. 
The ordering parameter is then that $\daone$ and $\dpone$, which capture local (mostly gyroscale) variation, are small.
This is the key step in this new derivation, introduced by Ref. \cite{Dimits_strongflow}: with the 
splitting in place, by avoiding a Taylor series expansion in gyroradius, standard drift-kinetic derivations
can be reworked to allow gyroscale fluctuations.

In order to capture most of the bulk plasma flow, the velocity shift $\ub$ is defined via
\begin{equation}
  \ub = \Eb \times \bb / B 
\end{equation}
with the electric field $ \Eb = - \nabla \phi - \partial \Ab / \partial t $, with all quantities defined
at guiding centre position $\Rb$.

We include most of the derivation here, although it is fairly routine, as an aid to the reader.
With these definitions, the Lagrangian can be expressed as
\begin{align*}
 \gamma = & \biggl[ \Ab(\Rb,t) + \rhob \cdot \nabla \Ab(\Rb,t) +  \delta_1 \Ab(\Rb,\rhob,t)  \\
             &  + B(\Rb) \rhob \times \bb(\Rb) + U \bb(\Rb) 
               + \ub \biggr]. d (\Rb +  \rhob)  \\
        - & \left[\frac{1}{2} (\vb+\ub)^2 + \phi(\Rb,t) + \rhob \cdot \nabla \phi(\Rb,t) + \delta_1 \phi(\Rb,\rhob,t) \right] dt.
\end{align*} 

We follow a recent approach by Brizard\cite{BrizardGyroangle} which avoids the appearance of gyrogauge
terms in the Lagrangian. The idea is that given a point $\Rb,\rhob,U,t$, we can define a gyroangle in the region
around this point, up to an arbitrary constant. Locally, we have $\rhob = \rhob(\theta,\rho,\Rb,t)$, with coordinate
dependence defined so that $d\rhob =  \hat{\rhob} d\rho + \bb \times \rhob d\theta + \rho [ d\Rb . \nabla_R + dt (\partial/\partial t) ] \hat{\rhob} $, with
\begin{equation}
\nabla_{\Rb,t} \hat{\rhob} =  - [ \nabla_{\Rb,t}  (\bb \bb) ] \hat{\rhob},
\end{equation}
which are orthogonal to the variations in $\rho$ and $\theta$.
This serves as a definition of the variation of the local gyroangle about some point $(\Rb,\rhob)$. It can be extended to
a definition of the gyroangle along a phase space trajectory $(\Rb(t),\rhob(t),t)$, which is what we need in order to define
variations of the trajectory, and therefore evaluate Euler-Lagrange equations. This definition of the gyroangle simplifies the
Lagrangian somewhat compared to the general case. However, it does not give rise to a consistent
global definition of $\theta$, as pointed out by Littlejohn\cite{LittleJohn81} (in their notation,
we have required that $e_1 .\nabla e_2 = 0$). The problem is that two trajectories starting and ending at the same points
$(\Rb,\rhob,U,t)$ may have different $\theta$ coordinates. Usually this is not a problem, because the aim is typically
to ignore the gyroangle, but one may keep track of $\hat{\rhob}$ if the physical angle is of interest.

It is often possible to define a global gyroangle $\theta(\Rb,\rhob,U,t)$, but 
additional terms are then present in the Lagrangian (gyrogauge terms). These terms effectively cancel drifts
which arise due to coordinate system rotation, so the actual particle motion is not sensitive to the choice of $\theta$. We feel
that these complications are best avoided.

So, inserting these coordinate definitions and field splittings in our Lagrangian, we have
\begin{align*}
 \gamma & = \left\{
             [ \Ab(\Rb,t) + U \bb(\Rb,t)  + \ub(\Rb,t) ].d\Rb  
             \right\} \\
      &  + \left\{
             [\Ab(\Rb,t) ].d\rhob
            +[\rhob \cdot \nabla \Ab(\Rb,t)].d\Rb
            +[ B(\Rb) \rhob \times \bb(\Rb,t)].d\Rb
          \right\} \\
      &    \biggl\{
              [\langle \Ab(\Rb,t) \rangle - \Ab(\Rb,t) + \delta_1 \Ab(\Rb,\rhob,t) ]
                 .d (\Rb + \rhob) \\
      &       [ U \bb(\Rb,t)  + \ub(\Rb,t) + \rhob \cdot \nabla \Ab(\Rb,t) + B(\Rb,t) \rhob \times \bb(\Rb,t) ]
                      .d\rhob
          \biggr\} + H
\end{align*}
All the tempero-spatially varying functions are defined at the point $\Rb,t$ and this is implicitly assumed in further expressions; there
is also an implicit dependence due to the constraint that $\rhob$ be perpendicular to $\bb$. We add 
a gauge $S = - \Ab . \rhob$ so terms in the second curly brace are rewritten
\begin{equation}
  \left\{ 
     - ( \ddt \Ab ) . \rhob dt + \rhob \cdot \nabla \Ab . d\Rb - d\Rb \cdot \nabla \Ab . \rhob
     +[ B \rhob \times \bb].d\Rb
  \right\}
\end{equation}
Which, apart from the first term, cancels on simplifying the expression using $ \Bb = \nabla \times A $. 

We also add a gauge $ S = -\rhob . \ub - (1/2) \rhob . \nabla A . \rhob $. The first term cancels a large
oscillation in gyroangle associated with bulk motion, while the second removes the electromagnetic 
vector potential $\Ab$ from certain expressions in favour of expressions involving the magnetic 
field $\Bb$. The terms of interest are then
\begin{align*}
 &  B \rhob \times \bb . d\rhob
  + \rhob \cdot \nabla \Ab .d\rhob - (1/2) d ( \rhob . \nabla \Ab . \rhob ) \\
=& \frac{1}{2} \rhob \cdot [ \nabla \Ab - (\nabla \Ab)^T ] \cdot d\rhob
   - (1/2) \rhob . \nabla ( [dR.\nabla + dt \ddt ] \nabla .\Ab ) . \rhob
  +  B \rhob \times \bb . d\rhob \\
=&  \frac{1}{2} B \rhob \times \bb . d\rhob - \frac{1}{2} \rhob . \nabla ( [dR.\nabla + dt \ddt ] \nabla .\Ab ) . \rhob \\
=&  \frac{1}{2} B \rho^2 d\theta 
   - \frac{1}{2} \rhob . \nabla ( [dR.\nabla + dt \ddt ] \nabla . \Ab ) . \rhob
\end{align*} 

We also use
\begin{equation}
 \rhob . (d \Rb . \nabla \nabla \Ab ) . \rhob =  [ \rhob \times (\rhob . \nabla \Bb) - (\rhob.\nabla)^2 : \Ab  ]. dR
\end{equation}

This results in a Lagrangian of the form
\begin{align*}
 \gamma & = [ \Ab + U \bb + \ub]. d\Rb + \frac{1}{2} B \rho^2 d\theta
    - \left[\frac{1}{2}(\ub^2+U^2) + B^2 \rho^2/2 + \phi \right] dt \\
   & + \epsilon \left\{ - \delta_1 \phi dt 
      - d(\ub + \delta_1 \Ab + U \bb ) . \rhob 
     +  \left[ \delta_1 \Ab - \frac{1}{2} (\rhob.\nabla)^2 : \Ab + \frac{1}{2} \rhob \times (\rhob . \nabla \Bb) \right] . d\Rb + \frac{1}{2} \rhob . \frac{\partial \nabla \Ab}{\partial t} . \rhob dt 
        \right\} \numberthis
\label{eq:guidecentrelag}
\end{align*}
where we have introduced a formal parameter $\epsilon$ (equal to unity) to denote that the set of terms in curly brackets is small.
For further manipulation, we will consider the Lagrangian to depend on the variable $\mu = B \rho^2/2$, so that the coefficient of $d \theta$ is simply $\mu$.

Note that at this point this expression is exact, and closed form.
All we have achieved is to split the Lagrangian into a large gyro-angle independent component and a small gyro-angle dependent component.
As well as terms associated with the extended ordering this differs from the guiding centre Lagrangian of Littlejohn
(actually the guiding centre Lagrangian is an intermediate expression not explicitly shown in that paper) 
due to the slightly different definition of the guiding centre transform.

\section{The gyrocentre transform}

Splitting the Lagrangian allows us to directly identify the lowest order motion in the system, which is a combination of simple gyration,
propagation along the field line, and advection with the $E \times B$ velocity. $\theta$ and $\mu$ form an action-angle pair of coordinates in
the `unperturbed' system, with $\mu$ a conserved quantitity; introducing a small perturbation to the system does not qualitatively modify the
dynamics, but somewhat modifies the quantity $\mu'$ that is conserved and its conjugate angle $\theta'$. We proceed, as is standard in gyrokinetic
analysis, to systematically find a new coordinate system (gyrocentre coordinates) where $\mu'$ is conserved by using the Lie transform technique.
This allows us to write a gyrocentre Lagrangian independent of gyroangle.
At first order, this is a straightforward extension of previous work\cite{Amil_GK}, and the lowest order coordinate perturbation is roughly
the sum of that required in the gyrokinetic\cite{Amil_GK} and drift kinetic\cite{LittleJohn83} case.

We briefly review the definitions of the transforms used here. A general coordinate transform $T$ is defined by the successive application of individual
transforms $T = .. T_3 T_2 T_1$. We define these as $ T_j(Z) = \exp \left( \epsilon^n \mathcal{L}_j \right) $; the transform may be considered to
arise via a displacement of the coordinates due to a flow $\mathcal{L}_j$ through the space (so, for example, a rotation would the produced by
a flow in the angular direction). Although there are simpler ways to define a near-unity coordinate transform, the Lie transform approach has a clear geometrical interpretation and useful algebraic properties, and is therefore more powerful. Applying this transform to the ordered Lagrangian
(eq. \ref{eq:guidecentrelag}) gives us a power series expression for the Lagrangian in the new coordinate system; we then proceed to manipulate
each order $n$ of this Lagrangian into the desired form by choosing $\mathcal{L}_n$ appropriately.


The action of the flow on scalars may be written  $\mathcal{L}_i (f) = g_i^{\mu} \partial f/\partial z^\mu$ in terms of
the generators $g^{\mu}_i$ which we wish to find. The transformation for the Lagrangian one-form
\begin{equation}
  (\mathcal{L}_i \gamma)_\mu = g_i^\sigma \left(\frac{\partial \gamma_\mu}{\partial z^{\sigma}}
  - \frac{\partial \gamma_\sigma}{\partial z^{\mu}} \right)
                            = g_i^\sigma \omega_{\sigma \mu}
\end{equation}
with the Poisson matrix defined as $\omega_{ij} = \partial_i \gamma_j - \partial_j \gamma_i$.
We now proceed to use these transforms to simplify the Lagrangian. With the definition $\gamma = \sum \epsilon^n \gamma_n $ we identify
$\gamma_1$ as the term in curly braces in eq. \ref{eq:guidecentrelag}, and $\gamma_0$ as the rest of the expression.
At lowest order, the Poisson matrix $\omega_{ij}^0$ derived from eq. \ref{eq:guidecentrelag}
is the same as for a simplified electrostatic formalism\cite{Amil_GK}, with the exception that the total (rather than just electrostatic)
electric field appears in the expression $\omega_{Rt}^0 = \Eb $.

\section{First order transform of the gyrocentre Lagrangian}

For the purpose of simplifying the notation, the first order terms can be written symbolically in the form
\begin{equation}
  \gamma_1 = -d \Mb  . \rhob + \datwo .  d\Rb  - \dptwo dt \label{eq:firstordergroup}
\end{equation}
with
\begin{equation}
  \dptwo = \delta_1 \phi - \frac{1}{2} \rhob . \frac{\partial \nabla \Ab}{\partial t} . \rhob,
\end{equation}
$\Mb = \ub + \delta_1 \Ab + U \bb $ and $\datwo = \delta_1 \Ab - \frac{1}{2} (\rhob.\nabla)^2 : \Ab + \frac{1}{2} \rhob \times (\rhob . \nabla \Bb)$. The first two terms in $\datwo$ almost cancel for long wavelength flows (which means they disappear entirely at this order in 
Cary and Littlejohn's analysis).

The transformed Lagrangian at this order 
\begin{align}
  \Gamma_1 = \gamma_1 - \mathcal{L}_1 \gamma_0 + dS_1.
  \label{eq:firstordersym}
\end{align}
The Lie transform is sufficiently general that we may set all the components of $\Gamma_1$ to zero except for
the time component; we remove any $\theta$-dependence in the time component but may be left with a secular part. 
  
We chose to group the terms in eq. \ref{eq:firstordergroup} in a way that maximises the possibility for
early simplification; each term is in the general form $ G dF $. Although the algebra will be presented later in a more
explicit fashion, we now explain the general principle that leads to this simplification. To find the 
generators at order $N$ arising due to a perturbation in $\gamma_1$ of the form $dF = (\partial_i F) dZ_i$ we evaluate
eq. \ref{eq:firstordersym} and find
\begin{equation}
  \omega_{ij}^0 g^{j}_N = \partial_i F
\end{equation}
for $i,j \neq 0$, which yields $g^{j}_N = (\omega_{ij}^0)^{-1} \partial_i F $. In the time-component, we have
terms $g^N_{j} \partial_j H_0 + \partial_t F $ contributing 
to $\mathcal{L}_N \gamma_0$.  Substituting, we find
$ \partial_j H_0 (\omega_{ij}^0)^{-1} \partial_i F = \partial_t F$. The lowest order trajectories
$\dot{Z}_{0i} = (\omega_{ij}^0)^{-1} (\partial_j H_0) $, so we can rewrite the  contribution in the time component
of $\mathcal{L}_N \gamma_0$ as $\dot{Z}_{0i} \partial_i F$ (with $i$ running from $0$). That is, the time-component
of the Lagrangian depends on the convective derivative of $F$ along the lowest order trajectory.
This is helpful because the ordering scheme requires that certain quantities vary slowly along
the lowest order trajectories. 

An explicit evaluation of eq. \ref{eq:firstordersym} yields
\begin{align}
  - \mathcal{L}_1 \gamma_0  = &    \gb_1^{\Rb}. \bb dU -  g_1^U \bb . d\Rb + \gb^{\Rb}_1 \times \Bb . d\Rb + g_1^\theta d\mu - g^\mu_1 d\theta \\ 
  & + ( \gb_{\Rb}^1 . [\partial \Ab/\partial t + \nabla \phi] +  U g_1^U +  g_1^\mu \Omega ) dt \nonumber \\
  & + \left\{ \gb^{\Rb}_1 \times \nabla \times ( U \bb + \ub ) . d\Rb
  + \gb^{\Rb}_1 . \nabla \left( \frac{1}{2} u^2 + \mu B \right) dt
    + \gb^{\Rb}_1 . \frac{\partial}{\partial t}( U \bb + \ub) dt \right\}
  \label{equation:eql1gamma0}
\end{align}
where the last term in curly brackets contains higher order contributions which will be promoted to the second order.  

The $U$ component of the Lagrangian gives 
\begin{equation}
  (\Gamma_1)_{U} = S_{1,U} + \gb_1^{\Rb} . \bb  
\end{equation}
so 
\begin{equation}
   \gb_1^{\Rb} . \bb = -S_{1,U}.  
\end{equation}
The $\Rb$ components yield
\begin{equation}
   (\Gamma_1)_{\Rb} = \gb_1^{\Rb} \times \Bb  -   \nabla \Mb . \rhob + \datwo + \nabla S_{1} - g_1^U \bb 
\end{equation}
by taking the cross product of this equation with $\bb$, we constrain the perpendicular component of $\gb_1^{\Rb}$.
The parallel component was found earlier, so overall we have
\begin{equation}
   \gb_1^{\Rb} =   \bb \times (\nabla \Mb . \rhob)/B - \bb \times (\datwo + \nabla S_1)/B
  - \bb S_{1,U}.
\end{equation}
Setting the parallel component of $(\Gamma_1)_{\Rb}=0$ requires 
\begin{equation}
  g_1^U  = - \bb . \nabla \Mb . \rhob + \bb . \datwo + \bb . \nabla S_1.
\end{equation}

\begin{equation}
  (\Gamma_1)_{\mu} = g_1^\theta + S_{1,\mu} - \frac{\partial \Mb}{\partial \mu } . \rhob 
\end{equation}
so
\begin{equation}
  g_1^{\theta} = -S_{1,\mu}  + \frac{\partial \Mb}{\partial \mu } . \rhob.
\end{equation}

Also
\begin{equation}
  (\Gamma_1)_{\theta} = -g_1^\mu + S_{1,\theta} - \frac{\partial \Mb}{\partial \theta } . \rhob,
\end{equation}
so
\begin{equation}
  g_1^{\mu}   = S_{1,\theta} - \frac{\partial \Mb}{\partial \theta} .\rhob.
\end{equation}

Using the definition of $\ub$, the time component can be written
\begin{equation}
  (\Gamma_1)_t = S_{1,t} + ( \Bb \times \gb_1^R . \ub - \bb . \gb_1^R \Eb_{\parallel} +  U g_1^U +  g_1^\mu \Omega ) - \dptwo - \frac{\partial \Mb}{\partial t } . \rhob
  \label{eq:gamma1t_init}
\end{equation}
and the term involving $\Eb_{\parallel}$, which is small, is promoted to second order. Substituting in the first order generators, we find
\begin{align}
(\Gamma_1)_t = - \frac{d \Mb}{dt_0} . \rhob + (\ub + U \bb) . \datwo - \dptwo  + \frac{d S}{dt_0}
\end{align}
with $ d/dt_0 = \partial/\partial t + U \partial/\partial R_{||} + \ub . \nabla + \Omega \partial/\partial \theta$, which is
the convective derivative along the lowest order solution for the trajectory.

Given that time variation is weaker in the moving frame, we will take the ansatz 
\begin{equation}
  (\partial_t + (\ub + U \bb ) . \nabla ) S_1  \sim \epsilon,
  \label{eq:s1approx}
\end{equation}
(implying that there is a term $d S/dt_0 - \partial S/\partial \theta_0$ promoted to second order)
so we can set the oscillatory part of $(\Gamma_1)_t$ to zero by solving
\begin{equation}
  S(\theta') = \frac{1}{\Omega} \int_{\theta_0}^{\theta'} d\theta \left( \frac{d \Mb}{d t_0} . \rhob - \left<\frac{d \Mb}{d t_0} . \rhob \right> - (\ub + U \bb) . (\datwo - \left<\datwo\right>)  + (\dptwo - \left<\dptwo\right>) \right)
\end{equation}
with $\theta_0$ set so that $\left<S\right>=0$.

The non-oscillatory components of $(\Gamma_1)_t$ are found from 
\begin{equation}
 \left< \dptwo \right> = \left< \phi(\Rb + \rho) - \phi(\Rb) - \rhob.\nabla \phi(\Rb) \right> - \left< \frac{1}{2} \rhob . \frac{\partial \nabla \Ab}{\partial t} . \rhob \right>
            = \left< \phi \right> - \phi -\frac{\rho^2}{4} \nabla_{\perp} . \frac{\partial \Ab}{\partial t},
\end{equation}
\begin{align*}
 \left<\datwo \right> &= \left<  \Ab(\Rb + \rho) - \Ab(\Rb) - \rhob.\nabla \Ab(\Rb) - \frac{1}{2} (\rhob.\nabla)^2 :: \Ab(R) + \rhob \times (\rhob . \nabla \Bb) \right> \\
           &= \left< \Ab \right> - \Ab - \frac{\rho^2}{4} \nabla_{\perp}^2 \Ab + \rho^2 \left< \bb (\hat{\rho} . \nabla \Bb) . \hat{\theta} - \hat{\theta} (\hat{\rho} . \nabla \Bb) . \bb \right> \\
           &= \left< \Ab \right> - \Ab - \frac{\rho^2}{4} \nabla_{\perp}^2 \Ab + \rho^2 \bb (\nabla \times \bb . \Bb ) - \rho^2 \bb \times \nabla B,
\end{align*}
and
\begin{align*}
 \left< \frac{d}{d t_0} \Mb . \rhob \right> = &\left< \frac{d}{d t_0} \daone . \rhob \right>  = - \Omega \rho \left< \daone . \theta \right> \\
         &          = \frac{\Omega}{2 \pi} \int \daone . d \rhob 
                   = \frac{\Omega}{2 \pi} \int (\Ab(\Rb + \rhob) - \rho.\nabla \Ab(\Rb) - A(\Rb)) . d\rhob \\
         &          = \frac{B}{2 \pi} \int \Bb( \Rb' ) . d\mathbf{S} - \frac{\rho^2}{2} B^2
\end{align*}
where the surface integral gives the magnetic flux through the gyroring.

So overall, we have
\begin{align*}
 \Gamma_1 = &\biggl( \frac{\rho^2}{2} B^2 - \frac{B}{2 \pi} \int \Bb( \Rb' ) . d\mathbf{S} 
   + (\ub + U \bb) . (\left< \Ab \right> - \Ab - \frac{\rho^2}{4} \nabla_{\perp}^2 \Ab) \\ 
    & + U \rho^2 (\nabla \times \bb . \Bb ) - \rho^2 \ub . \bb \times \nabla B 
   - \left< \phi \right> + \phi -\frac{\rho^2}{4} \nabla_{\perp} . \frac{\partial \Ab}{\partial t} \biggr) dt
\end{align*}
which can be interpretated in terms of additional kinetic energy due to drifts in the
direction of the lowest order motion, and departures of the gyroaveraged fields from the
local field values. This reduces to Littlejohn and Cary's gyrocentre Lagrangian when
the wavelength is ordered long.

\section{Second order transform of the gyrocentre Lagrangian }

The next order terms are found by systematially proceeding with the Lie transform. Since the perturbation
is not entirely in the Hamiltonian, this is somewhat more algebraically involved than many earlier
gyrokinetic theories. We have
\begin{equation}
  \Gamma_2 = \gamma_2 - \frac{1}{2} \mathcal{L}_1 (\gamma_1 + \Gamma_1) - \mathcal{L}_2 \gamma_0 + dS_2.
  \label{eq:secondorder}
\end{equation}

The relation
\begin{equation}
  L ( F  dG ) =  g_{\sigma} \frac{\partial G}{\partial z^\sigma} dF - g_{\sigma} \frac{\partial F}{\partial z^\sigma}  dG
  \label{eq:transformdef}
\end{equation}
which may be regarded as a definition of the transform, gives a result in the form $ K_1 dF + K_2 dG$.
Using earlier results, after solving for the generators to simplify the Lie-transformed Lagrangian,
the derivatives of $F$ and $G$ along zeroth order trajectories will then appear in the time component of
$\Gamma_2$, which, as in the first order calculation, helps to simplify the form of the Lagrangian
as certain terms may be neglected.


Substituting for $\gamma_1$ and using eq. \ref{eq:transformdef} we have 
\begin{align}
 \mathcal{L}_1 \Gamma_1 = & g^{\sigma}_1 \frac{\partial \rhob}{\partial z^\sigma} . d\Mb 
               - g^{\sigma}_1 \frac{\partial \Mb}{\partial z^\sigma} .  d\rhob
                - g^{\sigma}_1 \frac{\partial \datwo }{\partial z^\sigma} . d\Rb
               + d\Rb . \nabla \daone . \gb_{\Rb}^1
               + g^{\sigma}_1 \frac{\partial}{\partial z^{\sigma} } \dptwo dt.
                \label{eq:l1Gamma}
\end{align}
We also have
\begin{align}
  \mathcal{L}_1 \gamma_1 = -g_{\sigma}^1 \frac{\partial }{\partial z^\sigma} \left<\frac{d \daone}{d t_0} .\rhob \right> dt + g_{\sigma}^1 \frac{\partial \left<\datwo\right>}{\partial z^\sigma} . (\ub + U \bb) dt
  - g_{\sigma}^1 \frac{\partial}{\partial z^{\sigma} } \left<\dptwo\right> dt
  \label{eq:l1gamma}
\end{align}
and $\gamma_2$ is equal to the term in curly brackets in eq. \ref{equation:eql1gamma0}, plus terms from
eq. \ref{eq:gamma1t_init} and eq. \ref{eq:s1approx},  which were promoted
to second order. Substituting eqs. \ref{eq:l1gamma} and \ref{eq:l1Gamma} into eq. \ref{eq:secondorder}, and
then using the freedom of choice of the second order generators to set $(\Gamma)_{\sigma} = 0$ for $\sigma \neq 0$,
we find
\begin{align}
   (\Gamma_2)_t & = - \frac{g^{\sigma}_1}{2} \frac{\partial \rhob}{\partial z^\sigma} . \frac{d \Mb}{d t_0}
               + \frac{g^{\sigma}_1}{2} \frac{\partial \Mb}{\partial z^\sigma} . \frac{d\rhob}{d t_0}
               + \frac{g^{\sigma}_1}{2} \frac{\partial \datwot}{\partial z^\sigma} . (\ub + U \bb)
               - (\ub + U \bb) . \nabla \datwo . \frac{\gb_{\Rb}^1}{2} \nonumber \\
             &  - \frac{g^{\sigma}_1}{2} \frac{\partial}{\partial z^{\sigma} } \dptwot 
               + g^{\Rb}_1 . \frac{d}{d t_0} (\ub + U\bb) + \gb^{\Rb}_1 . \nabla (\mu B)
               + \frac{g^{\sigma}_1}{2} \frac{\partial }{\partial z^\sigma} \left<\frac{d \daone}{d t_0} .\rhob \right> \nonumber \\
             &  + \left(\frac{d}{dt_0} - \frac{\partial}{\partial \theta} \right) S_1
               - \bb . \gb_1^R \Eb_{\parallel}
               + \frac{d S_2}{d t_0}
               \label{eq:gamma2t}
\end{align}
where the terms with a tilde, $\datwot,\dptwot$ are the purely oscillatory components $\tilde a = a - \left<a\right>$.
The second order gauge $S_2$ is set to cancel the oscillatory part of $\Gamma_2$, and the remaining term that then needs to
be computed is the gyroaverage of eq. \ref{eq:gamma2t}.

Given that all the generators are known, we have an explicit form for the second order Lagrangian. This would usually
be regarded as an interim step in deriving the final expression for the Lagrangian, but it may actually be directly used
for numerical calculation. A relatively compact numerical evaluation should be possible, and many of the terms are zero.
Numerical integration over $\theta$ will be required, and standard integration techniques can perform this exactly for the terms
which have a polynomial dependence on $\rhob$.

Nevertheless, it is useful to compute the terms in this expression explicitly for purposes of interpretation and comparison with
other work: certain partial simplifications are possible, which are detailed in the appendices.


\section{A simplified Lagrangian for global microturbulence analysis}

We present a simplified version of the theory which has the correct small-scale
dynamics needed to model gyrokinetic microturbulence, but which has an appropriate minimal
non-perturbative model of large-scale dynamics. This simplification involves neglecting several second order
terms associated with the large scales; we justify this on the basis that the large scale Hamiltonian
is dominated by lower order terms. At short wavelength, the zeroth and first order Lagrangian have
weak dependence on the electromagnetic fields, and it is necessary to retain higher order terms
to define gyrokinetic Poisson and Ampere equations appropriate for microturbulence.

We also take advantage of the simplification of the Lagrangian in the case where the fluctuation $\daone = \bb \bb.\daone + O(\epsilon^2)$,
and as a consequence the local field strength varies weakly at the gyroscale. This is somewhat less restrictive
that the common approximation made in gyrokinetic codes that $\Ab = A_{\parallel} \bb$ which does not allow
magnetic compression even on long length scales.

The divergence terms associated with the $\gb_{\Rb}$ generator will be neglected here on the principle that they
lead to only a small modification of the gyrokinetic Ampere and Poisson equations as long as the spatial gradients
of the distribution function are small.

In this limit, the second order terms explicitly given in the appendices reduce to
\begin{equation}
<\Gamma_2> = -\frac{\Omega}{2} \ddmu \left( \frac{\partial S_1}{\partial \theta} \right)^2 dt,
\end{equation}
which is the same form seen in various earlier formulations. 

\section{A comment on the accuracy of weak-flow formalisms.}

Standard derivations of gyrokinetic theories\cite{Hahm_1988} initially ordered perturbed fields to be small,
but later the derivations were shown\cite{DimLoDub,Dimits_edge} to allow a more general ordering where the flows were weak, with $v_{E\times B} \ll v_{\perp}$,
where $v_{\perp}$ is the velocity associated with gyration.

This is almost universally interpreted as a condition on typical gyration
velocities, so the weak flow condition becomes $v_{E\times B} \ll v_{ti}$, with $v_{ti}$ a mean thermal speed. However, there are 
a fraction $\sim (v_{ti}/v_{E\times B})^2$ of particles for which the weak flow condition is not met, and for which the 
dervivation of gyrokinetic theory is not valid. Although this is only a small fraction of the particles, it isn't
immediately clear that we may proceed to use gyrokinetic theory as if all particles satisfied the ordering;  the power series
form of the Lagrangian is expected to diverge for this fraction, so the overall error even in collective behaviour would eventually
become unacceptable. 

Additionally, the condition that $k \rho \sim 1$ is broken in many cases: we are often interested in ions interacting with
wavelengths nearer the system size, or with short-wavelength turbulence driven by electrons.  

As an example of where the standard theory is invalid, we consider a particle in a homogeneous magnetic field $\Bb = \hat{z} B$ interacting
with a electrostatic potential made up of a large-amplitude background field, and a small short-wavelength fluctuation,
\begin{equation}
  \phi(\xb) = - x E_{x0} + \kappa \sin(k_x x).  
\end{equation}
For the weak-flow theory, we calculate the generators associated with the dominant long-wavelength field term as $g_{\Rb}=0$, $g_U=0$, 
\begin{equation}
  g_{\mu} = \frac{\partial S}{\partial \theta} \quad \mbox{and} \quad g_{\theta} = -\frac{\partial S}{\partial \mu}.
\end{equation}
with $S = (1/\Omega) \int d\theta \delta \phi = (\bb \times \rho) \nabla \phi / \Omega$. At lowest order the mapping
\begin{equation}
  \xb = \Rb + \rho \sim \bar{\Rb} + \bar{\rho} + g_{\mu} \frac{\partial \bar{\rho} }{\partial \bar{\mu} } + g_{\theta} \frac{\partial \bar{\rho} }{\partial \bar{\theta} }
  =\bar{\Rb} + \bar{\rho} + \frac{\partial S }{\partial \bar{\theta} }  \frac{\partial \bar{\rho} }{\partial \bar{\mu} } - \frac{\partial S }{\partial \bar{\mu} }\frac{\partial \bar{\rho} }{\partial \bar{\theta} }
  =\bar{\Rb} + \bar{\rho} + \bb \times \nabla_{\rho} S = \bar{\Rb} + \bar{\rho} + \nabla_{\perp} \phi / \Omega^2.
\end{equation}
in which the last term is simply the displacement $\db$ due to polarisation (the time integral of the polarisation drift). Even when $v_{E\times B} > v_{\perp}$, the weak-flow gyrocentre transform
correctly represents the polarisation displacement, suggesting that the theory may be appropriate even outside the regime where it has been derived.
However, a limitation of the weak flow theory is that it depends on the potential and its derivatives evaluated on the gyroring $\bar{\Rb} + \bar{\rho}$,
whereas the particle is actually located at the displaced position near $\bar{\Rb} + \bar{\rho} + \db$. The potential at the displaced position appears
in the weak-flow Lagrangian in the form of a Taylor series expansion $\phi(\hat{\Rb} + \db) = \sum_n (\db . \nabla)^n \phi / n! $.
In our example, we have $\phi(\Rb + \db) = \phi(\Rb) +  E_{x0}^2/\Omega + \kappa \sum_n (k_x E_{x0}/\Omega^2)^n / n!$,
which diverges for practical purposes (for the theories used in codes, $n \le 3$)
if $k_x v_{E\times B} / \Omega > 1$, and is a poor approximation at low order.
We should thus expect the weak flow formulation to give incorrect results when modelling systems where relatively strong flows exist
in conjunction with short-wavelength turbulence. 

Given that we have a theory more widely valid that weak-flow gyrokinetics, an obvious way to address such concerns is
to compare these theories directly. This provides a systematic way to justify the use of simpler theories, such as the usual
weak-flow theory.

\appendix

\section{Terms in the second order Hamiltonian}

We split up the calculation of $(\Gamma_2)_t$ in eq. \ref{eq:gamma2t} according to the index $\sigma$ of the generator
involved, with terms labelled $P_{\sigma}$; there is also term involving $S_1$, which gyroaverages to zero.
We will drop terms which have zero $\theta$ integral because these will be removed via a second order gauge function $S_2$ which we
will not need to explicitly calculate. Only terms of order $\epsilon^2$ or lower will be kept. 
To simplify the notation, we omit the lower index on the first order generators $g_1^{\sigma}$ and $S_1$ in this section.

\subsection{ Second order terms multiplied by $\mu$ and $\theta$ generators }

Take eq. \ref{eq:gamma2t}, and commute the $\theta$ and $\mu$ derivatives past $\dMdtz$ in first two terms (this commutation
is only valid at lowest order). We also use the definition of $S_1$, and find
\begin{align}
  2 P_{\mu} + 2 P_{\theta} = & - \Omega g_{\theta} \ddtheta \dsdtheta - \Omega g_{\mu} \ddmu \dsdtheta \notag 
    + g_{\theta} \rhob . \ddtheta \dMdtz + g_{\mu} \rhob . \ddmu \dMdtz \\
  & + g_{\theta} \drhodtz .\dMdtheta + g_{\mu} \drhodtz \dMdmu
\end{align}
At this point note that $\Mb$ appears with a $\mu$ or $\theta$ derivative; only $\daone$ contributes. We also have
$(d/dt_0) \daone \sim (\partial/\partial \Omega) \daone$ at this order. Theta derivatives
may be integrated by parts and we obtain
\begin{align}
  \mbox{RHS} = - \Omega g_{\theta} \ddtheta \dsdtheta - \Omega g_{\mu} \ddmu \dsdtheta
  + \Omega g_{\theta} \ddtheta \left(\rhob . \frac{\partial \daone}{\partial \theta} \right)
    + \Omega g_{\mu} \ddtheta \left(\rhob .\frac{\partial \daone}{\partial \mu} \right) \notag. \\
\end{align}
From the definitions of the generators, this expression depends entirely on $\daone$ and $S$.
This simplifies substantially in the case where $\daone \propto \bb$, in which case, we find that
\begin{align}
  \mbox{RHS} = -\Omega \frac{\partial}{\partial \mu} \left( \dsdtheta \right)^2.
\end{align}
As in earlier gyrokinetic derivations, the first order gauge function (and associated coordinate shift) cancels the $\theta$-dependent
effective potential in the first-order workings, but reappears at second order as the $\theta$-dependent displacements interact with $\theta$-dependent
fields (this is a ponderomotive-type effect). 

\subsection{ Second order terms multiplied by $U$ generator }
The derivative of $S$ along the field line is small, so $g_U$ simplifies and we have
\begin{align*}
  P_U = - \frac{g_U}{2} \bb . \frac{d \rhob}{d t_0}
  =  \frac{1}{2} [  \bb . \nabla \Mb . \rhob - \bb . \datwo ] \left[(U \bb + \ub).\nabla + \frac{\partial}{\partial t} \right] \rhob.\bb. \\
\end{align*}
As we have the derivative along the field line is $O(\epsilon)$, only the $O(1)$ terms of $M$ survive so this
\begin{align*}
  = -\frac{1}{2} [ \bb . \nabla ( U \bb + \ub ) . \rhob - \bb . \datwo ] \left[(U \bb + \ub).\nabla + \frac{\partial}{\partial t} \right] \bb.\rhob \\
  = -\frac{\rho^2}{4} \left( [U \bb + \ub].\nabla + \frac{\partial}{\partial t} \right) \bb . [ \bb . \nabla ( U \bb + \ub ) - \bb . \left<\datwo \rhob\right> ] 
\end{align*}
where we have used $\left< \mathbf{A} .\rhob \mathbf{C}. \rhob \right> =  \rho^2 \mathbf{A} . \mathbf{C}_{\perp} /2 $ when $A$ and $C$ are independent of $\theta$.
The long-wavelength, weak-flow component may be interpreted as the combined effect of the curvature of the field line and the FLR, which result in the effective
particle velocity differing from the gyrocentre velocity.

\subsection{ Second order terms multiplied by $\Rb$ generator }

For $\sigma=\Rb$ we have from eq. \ref{eq:gamma2t} (the term involving $E_{\parallel}$ for been dropped
as it has zero gyroaverage)
\begin{align}
&  P_{\Rb} = \frac{g_{\Rb}}{2} . \nabla \rhob . \frac{d \Mb}{d t_0}
  -  \frac{g_{\Rb}}{2} . \nabla \Mb . \frac{d \rhob}{d t_0} - \frac{g_{\Rb}}{2} . \nabla [ \datwot . (U \bb + \ub ) - \dptwo ] + (\ub + U \bb) . \nabla \datwot . \frac{g_{\Rb}}{2} \nonumber \\
 & - g_{\Rb}^1 . \frac{d}{d t_0} (\ub + U\bb) + \gb_{\Rb}^1 . \nabla (\mu B)
   - \frac{g_{\Rb}^1}{2} . \nabla \left<\frac{d \daonet}{d \theta} .\rhob \right>\\
   = & -g_{\Rb} . \nabla \left( \frac{\partial \rhob}{\partial \theta} . \daonet - \left< \frac{\partial \rhob}{\partial \theta} . \daonet \right> - \datwo . (U \bb + \ub ) + \dptwo  -\mu B \right) \nonumber \\
   &  - g_{\Rb}.\nabla  ( U \bb + \ub ) . \frac{\partial \rhob}{\partial \theta} + (\ub + U \bb) . \nabla \datwo . g_{\Rb}
     - g_{\Rb}^1 . \frac{d}{d t_0} (\ub + U\bb) 
\end{align}
In general this does not simplify substantially, but if $\daone . \rho= 0 $ then $\nabla . g_{\Rb}=0 $ and the first term may be written as a divergence: the divergence terms
can often be neglected for the purposes of deriving a quasineutrality equation as a consequence of the gradients in the distribution function being small.

\bibliography{gyrokin}

\begin{thebibliography}{9}
\expandafter\ifx\csname natexlab\endcsname\relax\def\natexlab#1{#1}\fi
\expandafter\ifx\csname bibnamefont\endcsname\relax
  \def\bibnamefont#1{#1}\fi
\expandafter\ifx\csname bibfnamefont\endcsname\relax
  \def\bibfnamefont#1{#1}\fi
\expandafter\ifx\csname citenamefont\endcsname\relax
  \def\citenamefont#1{#1}\fi
\expandafter\ifx\csname url\endcsname\relax
  \def\url#1{\texttt{#1}}\fi
\expandafter\ifx\csname urlprefix\endcsname\relax\def\urlprefix{URL }\fi
\providecommand{\bibinfo}[2]{#2}
\providecommand{\eprint}[2][]{\url{#2}}

\bibitem[{\citenamefont{Dimits}(2010)}]{Dimits_strongflow}
\bibinfo{author}{\bibfnamefont{A.~M.} \bibnamefont{Dimits}},
  \bibinfo{journal}{Physics of Plasmas} \textbf{\bibinfo{volume}{17}},
  \bibinfo{pages}{055901} (\bibinfo{year}{2010}).

\bibitem[{\citenamefont{Parra and Calvo}(2011)}]{Parra_calvo}
\bibinfo{author}{\bibfnamefont{F.~I.} \bibnamefont{Parra}} \bibnamefont{and}
  \bibinfo{author}{\bibfnamefont{I.}~\bibnamefont{Calvo}},
  \bibinfo{journal}{Plasma Physics and Controlled Fusion}
  \textbf{\bibinfo{volume}{53}}, \bibinfo{pages}{045001}
  (\bibinfo{year}{2011}).

\bibitem[{\citenamefont{Brizard and de~Guillebon}(2012)}]{BrizardGyroangle}
\bibinfo{author}{\bibfnamefont{A.~J.} \bibnamefont{Brizard}} \bibnamefont{and}
  \bibinfo{author}{\bibfnamefont{L.}~\bibnamefont{de~Guillebon}},
  \bibinfo{journal}{Physics of Plasmas} \textbf{\bibinfo{volume}{19}},
  \bibinfo{pages}{094701} (\bibinfo{year}{2012}).

\bibitem[{\citenamefont{Littlejohn}(1981)}]{LittleJohn81}
\bibinfo{author}{\bibfnamefont{R.~G.} \bibnamefont{Littlejohn}},
  \bibinfo{journal}{Physics of Fluids} pp. \bibinfo{pages}{1730--1749}
  (\bibinfo{year}{1981}).

\bibitem[{\citenamefont{Sharma and McMillan}(2015)}]{Amil_GK}
\bibinfo{author}{\bibfnamefont{A.~Y.} \bibnamefont{Sharma}} \bibnamefont{and}
  \bibinfo{author}{\bibfnamefont{B.~F.} \bibnamefont{McMillan}},
  \bibinfo{journal}{Physics of Plasmas} \textbf{\bibinfo{volume}{22}},
  \bibinfo{eid}{032510} (\bibinfo{year}{2015}).

\bibitem[{\citenamefont{Littlejohn}(1983)}]{LittleJohn83}
\bibinfo{author}{\bibfnamefont{R.~G.} \bibnamefont{Littlejohn}},
  \bibinfo{journal}{Journal of Plasma Physics} \textbf{\bibinfo{volume}{29}},
  \bibinfo{pages}{111} (\bibinfo{year}{1983}).

\bibitem[{\citenamefont{Hahm}(1988)}]{Hahm_1988}
\bibinfo{author}{\bibfnamefont{T.}~\bibnamefont{Hahm}},
  \bibinfo{journal}{Physics of Fluids} \textbf{\bibinfo{volume}{31}},
  \bibinfo{pages}{2670} (\bibinfo{year}{1988}).

\bibitem[{\citenamefont{Dimits et~al.}(1992)\citenamefont{Dimits, LoDestro, and
  Dubin}}]{DimLoDub}
\bibinfo{author}{\bibfnamefont{A.~M.} \bibnamefont{Dimits}},
  \bibinfo{author}{\bibfnamefont{L.~L.} \bibnamefont{LoDestro}},
  \bibnamefont{and} \bibinfo{author}{\bibfnamefont{D.~H.~E.}
  \bibnamefont{Dubin}}, \bibinfo{journal}{Physics of Fluids B}
  \textbf{\bibinfo{volume}{4}}, \bibinfo{pages}{274} (\bibinfo{year}{1992}).

\bibitem[{\citenamefont{Dimits}(2012)}]{Dimits_edge}
\bibinfo{author}{\bibfnamefont{A.~M.} \bibnamefont{Dimits}},
  \bibinfo{journal}{Physics of Plasmas} \textbf{\bibinfo{volume}{19}},
  \bibinfo{eid}{022504} (\bibinfo{year}{2012}).

\end{thebibliography}

\end{document}